\documentclass[preprint]{emulateapj}

\begin{document}

\title{Search for the Return of Activity in Active Asteroid 176P/LINEAR}

\author{
%Contributors
Henry H.\ Hsieh\altaffilmark{1,2,a},
Larry Denneau\altaffilmark{2},~%UH - help with MOPS
Alan Fitzsimmons\altaffilmark{3},~%QUB - on VLT proposal
Olivier R.\ Hainaut\altaffilmark{4},~%ESO - ESO large program
Masateru Ishiguro\altaffilmark{5},~%SNU - obtained subaru data
Robert Jedicke\altaffilmark{2},~%UH - PS1 support
Heather M.\ Kaluna\altaffilmark{2},~%UH - obtained uh88 data + calibration fields
Jacqueline V.\ Keane\altaffilmark{2},~%UH - obtained uh88 data
Jan Kleyna\altaffilmark{2},~%UH - obtained uh88 data
Pedro Lacerda\altaffilmark{3,6,b},~%QUB - on VLT proposal
Eric M.\ MacLennan\altaffilmark{2,7},~%NAU - helped with 2010 uh88 176p data reduction
Karen J.\ Meech\altaffilmark{2},~%UH - supervises heather
Nick A.\ Moskovitz\altaffilmark{8},~% - obtained Dupont data
Timm Riesen\altaffilmark{2},~%UH - obtained 2010 uh88 data
Eva Schunova\altaffilmark{2},~%UH - helped with keck observing
Colin Snodgrass\altaffilmark{6},~%MPS - on VLT proposal
Chadwick A.\ Trujillo\altaffilmark{9},~%Gemini - helped with 2010 gemini 176p observations
Laurie Urban\altaffilmark{2},~%UH - obtained uh88 data
Peter Vere{\v s}\altaffilmark{2},~%UH - helped with PS1 photometry
Richard J.\ Wainscoat\altaffilmark{2},~%UH - PS1 support
Bin Yang\altaffilmark{2,10},~%UH - obtained 2010 uh88 data
}
%\affil{
\altaffiltext{1}{Institute for Astronomy and Astrophysics, Academia Sinica, No.\ 1, Sec.\ 4, Roosevelt Rd., Taipei 10617, Taiwan}
\altaffiltext{2}{Institute for Astronomy, University of Hawaii, 2680 Woodlawn Drive, Honolulu, HI 96822, USA}
\altaffiltext{3}{Astrophysics Research Centre, Queens University Belfast, Belfast BT7 1NN, United Kingdom}
\altaffiltext{4}{European Southern Observatory, Karl-Schwarzschild-Stra\ss e 2, D-85748 Garching bei MŸnchen, Germany}
\altaffiltext{5}{Department of Physics and Astronomy, Seoul National University, 599 Gwanak-ro, Gwanak, Seoul 151-742, Republic of Korea}
\altaffiltext{6}{Max-Planck-Institut f\"ur Sonnensystemforschung, Max-Planck-Str.\ 2, 37191 Katlenburg-Lindau, Germany}
\altaffiltext{7}{Department of Earth and Planetary Sciences, University of Tennessee, 306 EPS Building, 1412 Circle Drive, Knoxville, TN 37996, USA}
\altaffiltext{8}{Department of Earth, Atmospheric \& Planetary Sciences, Massachusetts Institute of Technology, 77 Massachusetts Ave., Cambridge, MA 02139, USA}
\altaffiltext{9}{Gemini Observatory, Northern Operations Center, 670 N.\ AÔohoku Place, Hilo, HI 96720, USA}
\altaffiltext{10}{European Southern Observatory, Alonso de C\'ordova 3107, Vitacura, Casilla 19001, Santiago de Chile, Chile}
%\altaffiltext{3}{Department of Astronomy, Faculty of Mathematics, University of Belgrade, Studentski trg 16, 11000 Belgrade, Serbia}
%\altaffiltext{4}{Las Cumbres Observatory Global Telescope Network, Inc., 6740 Cortona Dr.\ Suite 102, Santa Barbara, CA 93117 USA}
%\altaffiltext{5}{Centre for Astrophysics and Planetary Science, The University of Kent, Canterbury CT2 7NH, United Kingdom}
%\altaffiltext{6}{Gemini Observatory, Northern Operations Center, 670 N.\ AÔohoku Place, Hilo, HI 96720, USA}
%\altaffiltext{7}{Institute of Astronomy, National Central University, 300 Jhongda Rd, Jhongli 32001, Taiwan}
%\altaffiltext{8}{Department of Physics, P.O. Box 64, 00014 University of Helsinki, Finland}
%\altaffiltext{9}{Planetary Science Institute, 1700 East Fort Lowell, Suite 106, Tucson, AZ 85719}
%\altaffiltext{10}{Dipartimento di Matematica, Universit{\`a} di Pisa, Largo Pontecorvo 5, 56127 Pisa, Italy}
%\altaffiltext{11}{Planetary Exploration Group, Space Department, Johns Hopkins University Applied Physics Laboratory, Laurel, MD 20723, USA}
%\altaffiltext{12}{Department of Astronomy, University of Maryland, College Park, MD 20742, USA}
%\altaffiltext{13}{Department of Physics, University of Central Florida, 4000 Central Florida Blvd., Orlando, FL 32816, USA}
%\altaffiltext{14}{Indian Institute of Astrophysics, CREST Campus, Block-II, Koramangala, Sarjapur Road, Bangalore 560034, India}
%\altaffiltext{15}{Department of Astrophysical Sciences, Peyton Hall, Princeton University, Princeton, 08544, USA}
\altaffiltext{a}{Hubble Fellow}
\altaffiltext{b}{Michael West Fellow}
%}
\email{hsieh@ifa.hawaii.edu}

\slugcomment{Updated, 2014-01-03}

\begin{abstract}
We present the results of a search for the reactivation of active asteroid 176P/LINEAR during its 2011 perihelion passage using deep optical observations obtained before, during, and after that perihelion passage.  Deep composite images of 176P constructed from data obtained between June 2011 and December 2011 show no visible signs of activity, while photometric measurements of the object during this period also show no significant brightness enhancements similar to that observed for 176P between November 2005 and December 2005 when it was previously observed to be active.  An azimuthal search for dust emission likewise reveals no evidence for directed emission (i.e., a tail, as was previously observed for 176P), while a one-dimensional surface brightness profile analysis shows no indication of a spherically symmetric coma at any time in 2011.  We conclude that 176P did not in fact exhibit activity in 2011, at least not on the level on which it exhibited activity in 2005, and suggest that this could be due to the devolatization or mantling of the active site responsible for its activity in 2005.
\end{abstract}


\newpage

\section{INTRODUCTION}\label{section_introduction}

During the past several years, a number of objects orbiting in the main asteroid belt exhibiting comet-like activity have been discovered.  The first of these objects, 133P/Elst-Pizarro, discovered in 1996 \citep{els96}, was originally suspected of exhibiting comet-like activity as a result of an impact on its surface by another asteroid \citep[e.g.,][]{tot00}.  Observations of repeated activity during subsequent perihelion passages \citep{hsi04,hsi10,low05,hsi13}, however, provide strong evidence against this initial explanation, and indicate instead that 133P's activity is most likely cometary in nature, i.e., driven by the sublimation of volatile ice \citep{hsi04}.  The subsequent discoveries in 2005 of two more such objects, 238P/Read and 176P/LINEAR \citep{rea05,hsi06a}, led to the designation of these objects as a new class of comets, the main-belt comets \citep[MBCs;][]{hsi06}.

While additional MBCs continue to be discovered \citep[259P/Garradd, P/2010 R2 (La Sagra), P/2006 VW$_{139}$, P/2012 T1 (PANSTARRS), P/2013 R3 (Catalina-PANSTARRS);][]{gar08,nom10,hsi11c,wai12,hil13}, objects that exhibit comet-like dust emission that does in fact appear to result from impacts or rotational disruption \citep[P/2010 A2 (LINEAR), (596) Scheila, P/2012 F5 (Gibbs), P/2013 P5 (PANSTARRS);][]{jew10,jew11,jew13,sno10,bod11,ish11,ste12} have also been discovered.  Distinguishing sublimation-driven cometary activity from dust emission resulting from impacts, rotational disruption, or any of several other potential dust ejection mechanisms unrelated to sublimation \citep{jew12} is not a simple task, however.  Direct spectroscopic detection of sublimation products in any MBC has proven extremely difficult to achieve, eluding all attempts to directly confirm the presence of sublimation to date \citep{jew09,hsi12b,hsi12c,hsi13a,dev12,oro13}.  These non-detections are likely due to the faintness of the objects (with typical V-band magnitudes of $m_V > 18$) and low outgassing rates \citep[cf.][]{hsi04}, and do not actually rule out the presence of sublimation products, but rather simply indicate that they fall below the detection limits of the discovery attempts made to date.

Due to the diversity of dynamically asteroidal objects exhibiting apparent cometary activity, as well as the diversity of opinions on what the defining observational, physical, and dynamical properties of asteroids and comets are, a variety of names have been suggested for such objects, including ``activated asteroids'' \citep[cf.][]{hsi04}, ``main-belt comets'' \citep{hsi06}, ``active asteroids'' \citep{jew12}, and ``active main-belt objects'' \citep{bau12}.  In this work, we will use the term ``main-belt comet'' to refer to objects that occupy stable main-belt asteroid orbits and exhibit comet-like activity driven by the sublimation of volatile material, and the term ``disrupted asteroid'' to refer to objects that exhibit dust emission due to recent impacts \cite[cf.][]{hsi12a}.  Meanwhile, we will use the term ``active asteroid'' to describe any object with a dynamically asteroidal orbit that exhibits comet-like activity, whether the source of that activity is sublimation, a recent impact, or unclear or unknown.

Also known as asteroid (118401) 1999 RE$_{70}$, 176P was discovered to exhibit comet-like dust emission on 2005 November 26 by the Gemini North telescope on Mauna Kea in Hawaii \citep{hsi06a}, shortly after it passed perihelion on 2005 October 18.  Follow-up observations confirmed the presence of this activity, verified its persistence as late as 2005 December 29, and finally documented its disappearance by 2006 February 3 \citep{hsi11a}.  Anticipating the reappearance of activity following its next perihelion passage on 2011 July 1, \citet{dev12} performed a spectroscopic search for emission indicative of the presence of H$_2$O using the Heterodyne Instrument for the Far Infrared onboard the {\it Herschel} Space Observatory on 2011 August 8.  No H$_2$O line emission was detected suggesting that any sublimation that may have been present was below the detection limits of the observations conducted, but also that the comet was likely less active during the HIFI observations than it was when observed in 2005 \citep{dev12}.

In this work, we present deep optical observations of 176P obtained before, after, and during that 2011 perihelion passage in order to determine whether or not active dust emission did in fact resume during this period, where the confirmed resumption of activity would suggest very strongly that this active asteroid's activity is indeed sublimation-driven and a non-detection of activity would cast significant doubt on that conclusion.

\section{OBSERVATIONS}\label{section_observations}

\begin{figure}
\plotone{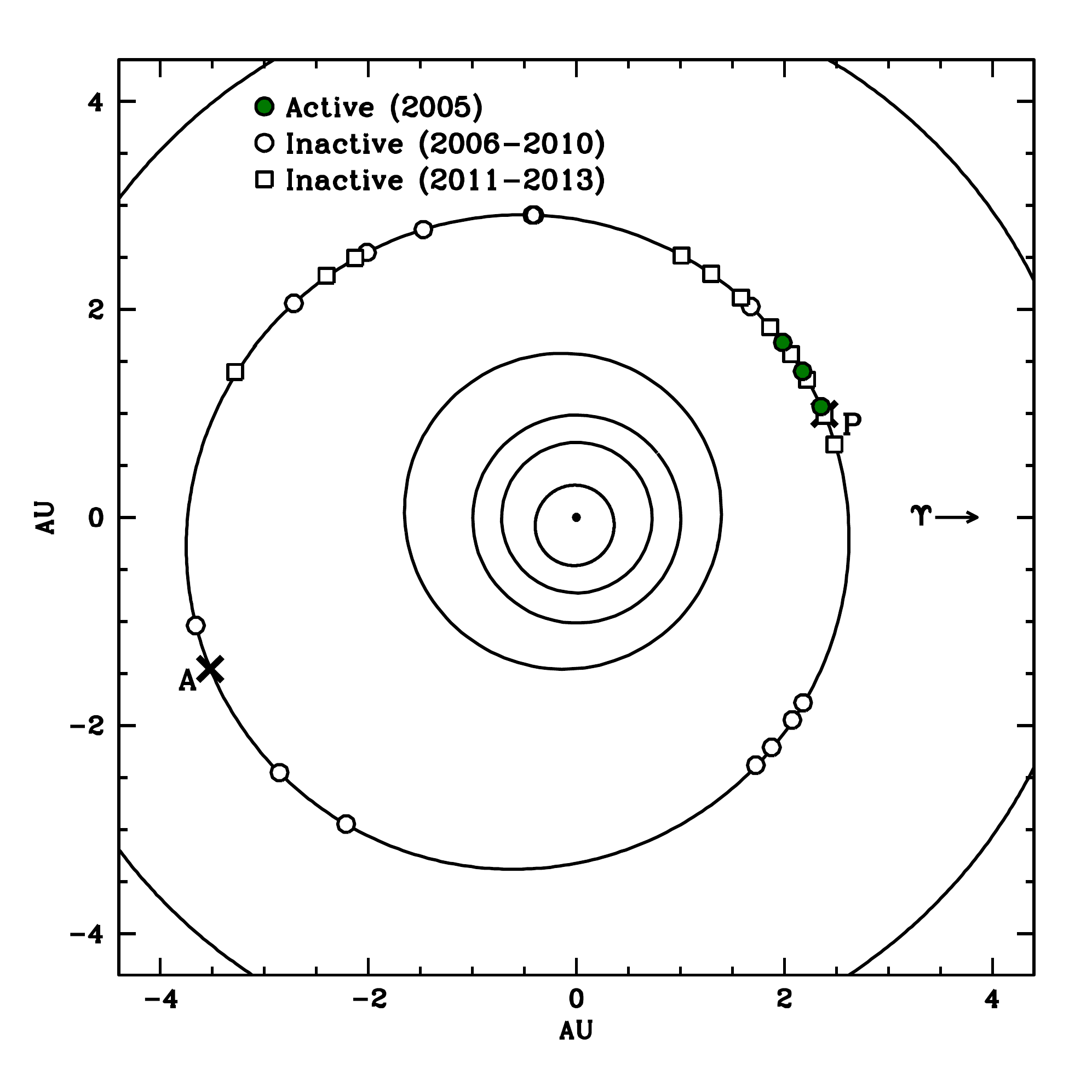}
\caption{\small 
Plot of orbital positions of 176P/LINEAR during observations detailed in Tables~\ref{table_obslog_new} and \ref{table_obslog_old}, where green circles mark observations obtained in 2005 when 176P was observed to be active, open circles mark observations obtained between 2006 and 2010 when 176P was observed to be inactive, and open squares mark observations obtained during and following 176P's 2011 perihelion passage when it was also observed to remain inactive.
}
\label{fig_orbitplot}
\end{figure}

Observations of 176P were obtained using multiple telescopes from 2010 through 2013, covering a substantial part of object's orbit during and after its 2011 July 1 perihelion passage.  In particular, our observations completely overlap the true anomaly range of the observations reported by \citet[][]{hsi11a}, allowing us to directly compare the object's active behavior over identical orbit arcs (Figure~\ref{fig_orbitplot}).

Facilities used include
the University of Hawaii (UH) 2.2~m,
the 8.2~m Subaru,
the 8.1~m Gemini North (Program GN-2011B-Q-17),
and the 10~m Keck I telescopes on Mauna Kea,
the 1.8~m Pan-STARRS1 (PS1) survey telescope on Haleakala,
the 2.5~m Ir\'en\'ee du Pont telescope at Las Campanas,
the 3.54~m New Technology Telescope (NTT; Programs 184.C-1143 and 085.C-0363(A)) operated by the European Southern Observatory (ESO) at La Silla, and
the 8.2~m Very Large Telescope (VLT; Program 086.C-0939(A)) operated by the European Southern Observatory (ESO) at Paranal.

We employed
a $2048\times2048$~pixel Textronix CCD and Kron-Cousins filters for UH observations,
the Subaru Prime Focus Camera, or Suprime-Cam \citep{miy02}, for Subaru observations,
the Gemini Multi-Object Spectrograph, or GMOS \citep{hoo04}, for Gemini observations,
the Low-Resolution Imaging Spectrometer, or LRIS \citep{oke95}, for Keck observations,
a 1.4 gigapixel camera consisting of a mosaic of 60 orthogonal transfer arrays, each consisting of 64 $590\times598$~pixel CCDs, for PS1 observations,
the ESO Faint Object Spectrograph and Camera, or EFOSC2 \citep{buz84}, for NTT observations,
and the visual and near-ultraviolet Focal Reducer and low dispersion Spectrograph, or FORS2 \citep{app98}, for VLT observations.

Gemini observations utilized a Sloan Digital Sky Survey (SDSS) $r'$-band filter, while PS1 observations were obtained using a filter designated $r_{\rm P1}$ that closely approximates the SDSS $r'$-band filter \citep{ton12}.  All other observations were conducted using Kron-Cousins $R$-band filters.  All observations were conducted using non-sidereal telescope tracking at the apparent rate and direction of the motion of 176P on the sky, except for PS1 observations which were conducted using sidereal tracking.
  
%Follow-up imaging was performed using the 2.0~m Faulkes Telescope North (FTN) on Haleakala and Faulkes Telescope South (FTS) at Siding Spring, the University of Hawaii (UH) 2.2~m telescope on Mauna Kea, the 1.8~m Perkins Telescope (PT) at Lowell Observatory, the 2.0~m Himalayan Chandra Telescope (HCT) at the Indian Astronomical Observatory on Mt.\ Saraswati, the 4.2~m William Herschel Telescope (WHT; Program SW2011b20) at La Palma, the 3.54~m New Technology Telescope (NTT; Program 185.C-1033(K)) operated by the European Southern Observatory (ESO) at La Silla, and the Lulin One-meter Telescope (LOT) in Taiwan.
%We employed $4096\times4096$~pixel Fairchild CCDs and either SDSS or Bessell filters for Faulkes observations, 
%a $2048\times2048$~pixel Textronix CCD and Kron-Cousins filters for UH observations,
%the Perkins ReImaging System (PRISM) and Kron-Cousins filters for Perkins observations,
%a $2048\times2048$~pixel E2V CCD and Bessell filters for HCT observations,
%the WHT's auxiliary-port camera \citep[ACAM;][]{ben08} and SDSS filters on the WHT, 
%the ESO Faint Object Spectrograph and Camera \citep[EFOSC2;][]{buz84} and Bessell filters on the NTT, 
%and a VersArray:1300B CCD \citep{kin05} and Bessell-like filters on the LOT.

We performed standard bias subtraction and flat-field reduction (using dithered images of the twilight sky) for all data, except those from PS1, using Image Reduction and Analysis Facility (IRAF) software \citep{tod86,tod93}.  PS1 data were reduced using the system's Image Processing Pipeline \citep[IPP;][]{mag06}.  Photometry of \citet{lan92} standard stars and field stars was performed by measuring net fluxes within circular apertures, with background sampled from surrounding circular annuli.  For Gemini, VLT, Subaru, and PS1 data, for which Landolt standards were not available, and other data obtained under non-photometric conditions, absolute calibration was accomplished using SDSS field star magnitudes \citep{aih11}.  Conversion of $r'$-band Gemini and PS1 photometry to $R$-band was accomplished using transformations derived by \citet{ton12} and by R.\ Lupton ({\tt http://www.sdss.org/}).  Comet photometry was performed using circular apertures with varying radii depending on the nightly seeing, where background statistics were measured in nearby, but non-adjacent, regions of blank sky to avoid dust contamination from the comet.  At least five field stars in all comet images were also measured to correct for extinction variation during each night.

\begin{figure}
\plotone{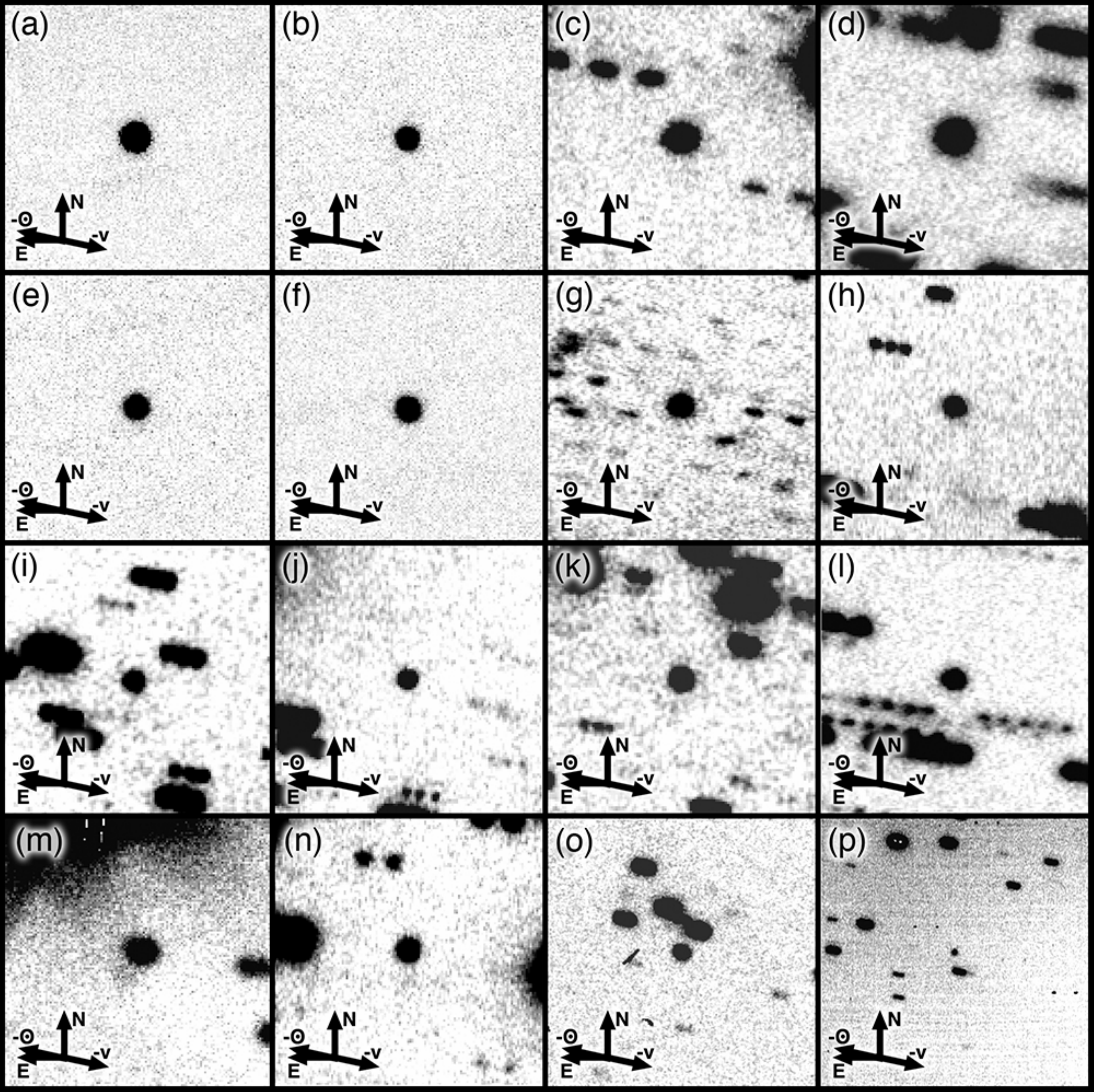}
\caption{\small 
Composite images of 176P from data obtained during 2010 observations detailed in Table~\ref{table_obslog_new} during which the object was expected to be inactive.  Each image is $0\farcm5\times0\farcm5$ with 176P at the center, with arrows indicating north (N), east (E), the negative heliocentric velocity vector ($-v$), and the projection of the antisolar vector on the sky ($-\odot$).  Images shown correspond to observations obtained on (a) 2010 August 5, and (b) 2010 August 6, both with the NTT, (c) 2010 August 11, and (d) 2010 August 13, both with the UH 2.2~m telescope, (e) 2010 August 14, and (f) 2010 August 15, both with the NTT, (g) 2010 August 16 with the UH 2.2~m, (h) 2010 August 28, (i) 2010 August 29, (j) 2010 August 30, and (k) 2010 August 31, all with the du Pont telescope, (l) 2010 September 1 with the UH 2.2~m, (m) 2010 September 4, and (n) 2010 September 5, both with the NTT, (o) 2010 October 5 with Keck I, and (p) 2010 October 23 with the VLT.
}
\label{fig_images_1}
\end{figure}

\begin{figure}
\plotone{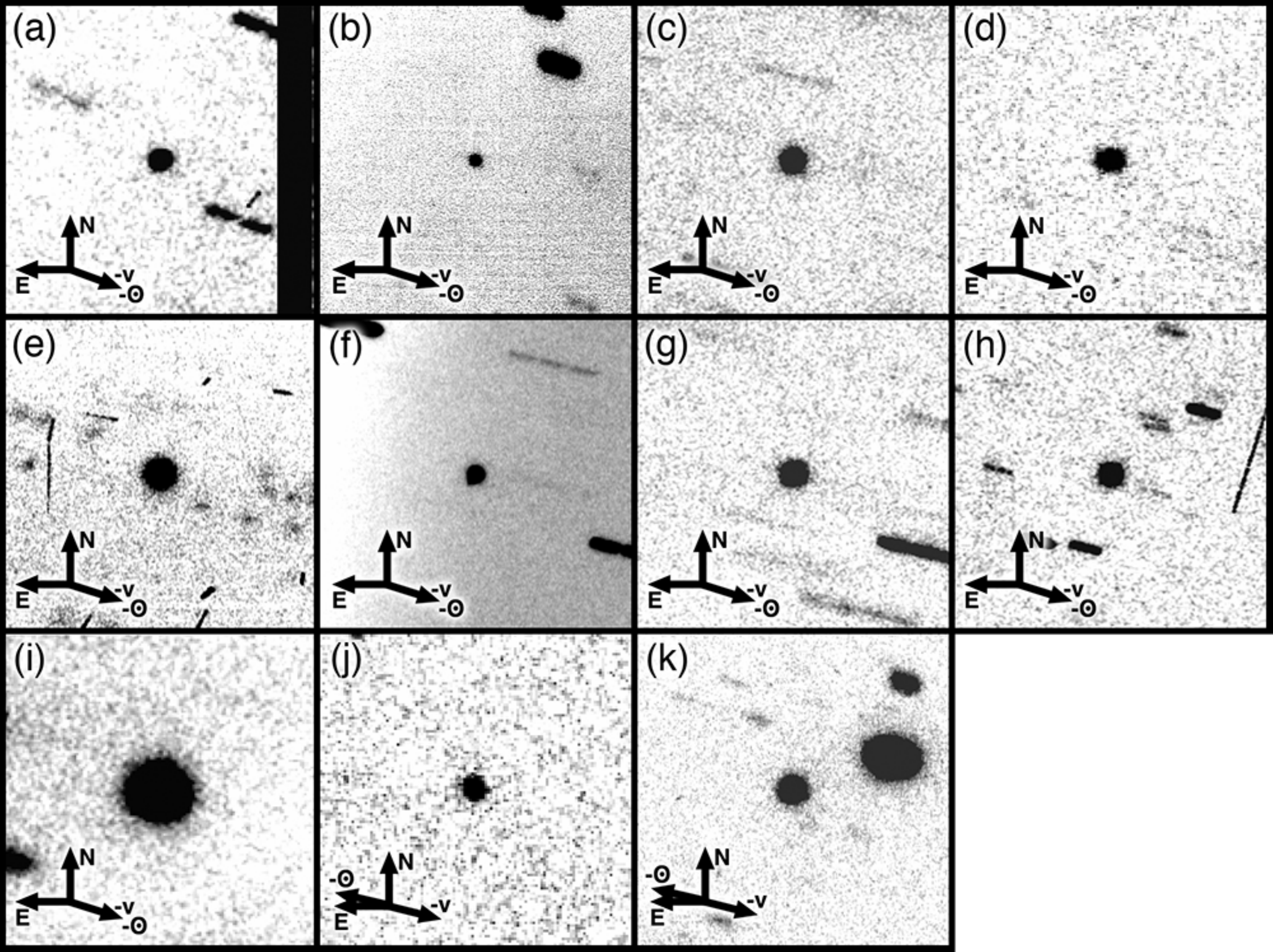}
\caption{\small 
Composite images of 176P from data obtained during 2011 observations detailed in Table~\ref{table_obslog_new} during which the object was expected to potentially become active.  Each image is $0\farcm5\times0\farcm5$ with 176P at the center, with arrows indicating north (N), east (E), the negative heliocentric velocity vector ($-v$), and the projection of the antisolar vector on the sky ($-\odot$).  Images shown correspond to observations obtained on (a) 2011 June 6 with Subaru, (b) 2011 July 1 with the VLT, (c) 2011 August 2 with Gemini, (d) 2011 August 4 with the UH 2.2~m telescope, (e) 2011 August 26 with Keck I, (f) 2011 August 28, (g) 2011 August 29, and (h) 2011 September 25, all with Gemini, (i) 2011 October 30 with the UH 2.2~m, (j) 2011 December 1 with PS1, and (k) 2011 December 31 with Gemini.
}
\label{fig_images_2}
\end{figure}

\begin{figure}
\plotone{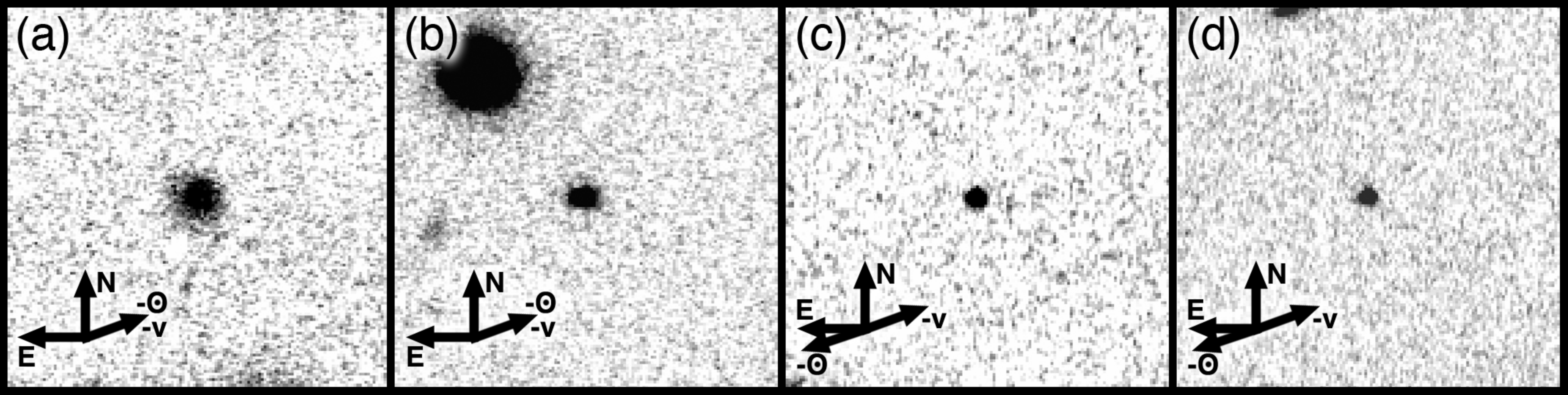}
\caption{\small 
Composite images of 176P from data obtained during 2012 and 2013 observations detailed in Table~\ref{table_obslog_new} during which the object was expected to be inactive.  Each image is $0\farcm5\times0\farcm5$ with 176P at the center, with arrows indicating north (N), east (E), the negative heliocentric velocity vector ($-v$), and the projection of the antisolar vector on the sky ($-\odot$).  Images shown correspond to observations obtained on (a) 2012 November 13, (b) 2012 December 18, (c) 2013 May 12, and (d) 2013 May 13, all with the UH 2.2~m telescope.  No visible activity is observed.
}
\label{fig_images_3}
\end{figure}

To maximize signal-to-noise ratios, we construct composite images of the object for each night of data by shifting and aligning individual images on the object's photocenter using linear interpolation and then adding them together.  The resulting composite images are shown in Figures~\ref{fig_images_1} (showing data obtained prior to the object's 2011 perihelion passage), \ref{fig_images_2} (showing data obtained during and shortly after perihelion), and \ref{fig_images_3} (showing data obtained well after perihelion).  We note that no visible cometary activity is immediately evident in any of these images.

\section{RESULTS AND ANALYSIS}\label{section_results}

\subsection{Photometric Analysis}\label{section_photanalysis}

One quantitative method for searching for low-level cometary activity is by using photometric analysis to investigate whether an object's brightness deviates from what is expected based on prior observations of the object when it was believed to be inactive.  An increase in brightness for an object that is otherwise stellar in appearance could indicate the presence of unresolved coma surrounding the object.  This technique was used to discover activity in 95P/(2060) Chiron \citep{bus88,tho88,mee89,har90}.

\begin{figure}
\plotone{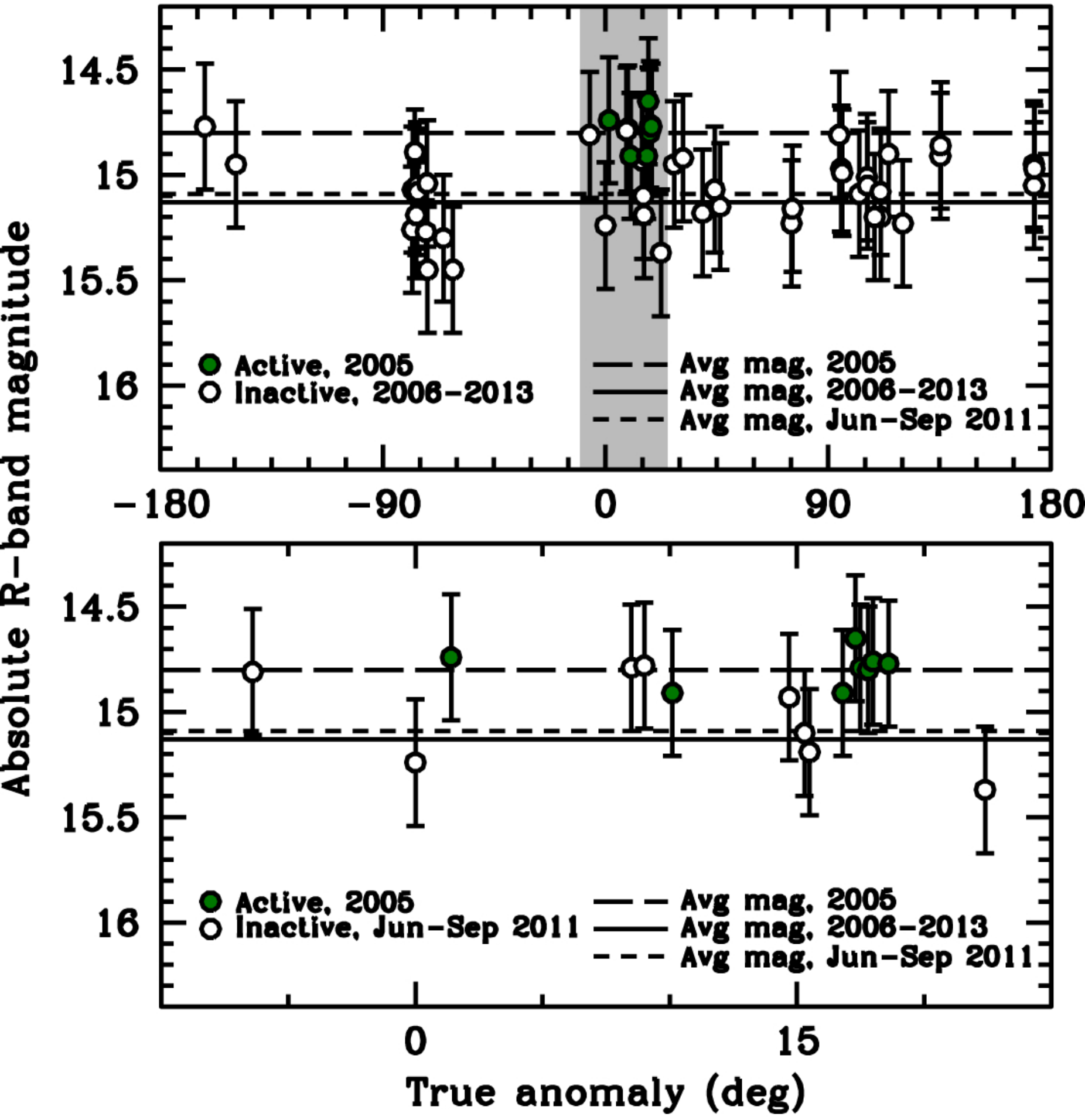}
\caption{\small 
Plots of absolute magnitudes measured for 176P/LINEAR as a function of true anomaly (a) over the object's full orbit, and (b) over a section of the object's orbit near perihelion over which it was observed to be active in 2005 and was expected to be active in 2011.  Observations obtained while the object was observed to be active in 2005 are marked by solid green circles.  Observations obtained while 176P appeared inactive are marked with open circles.  A gray shaded region highlights the true anomaly range over which activity was observed in 2005 and expected in 2011 (approximately between June and September 2011).  The average absolute magnitude of data obtained when 176P was active in 2005 is indicated by a large-dashed line in each panel while the average absolute magnitude of data obtained when 176P was expected to be active in 2011 is indicated by a short-dashed line in each panel.  The average magnitude of all data obtained from 2006 to 2013 when 176P appeared inactive is indicated by a solid line in each panel.
}
\label{fig_absmag}
\end{figure}

The inactive photometric behavior of 176P has been previously established by \citet{hsi11a} who derived best-fit IAU phase function parameters of $H=15.10\pm0.05$~mag and $G=0.15\pm0.10$.  Using $G=0.15\pm0.10$, we can then compute the equivalent absolute magnitudes (at heliocentric and geocentric distances of $R=\Delta=1$~AU and a solar phase angle of $\alpha=0^{\circ}$) for all of our new observations of 176P (Table~\ref{table_obslog_new}), and then plot these data as a function of true anomaly (Figure~\ref{fig_absmag}).  For comparison, we also plot previously reported photometric data for 176P from \citet{hsi11a} (shown in Table~\ref{table_obslog_old}).

As is evident from Figure~\ref{fig_absmag}, searching for activity in a small body in this way is complicated by rotational brightness variations.  \citet{hsi11a} found a rotational period for 176P of $P_{\rm rot}=22.23\pm0.01$~hr and a peak-to-trough photometric range of $\Delta m \sim 0.7$~mag (although this photometric range may vary at different observational epochs depending on aspect angle).  Thus, a significant amount of the scatter seen in Figure~\ref{fig_absmag} is likely due to rotational variations, and detecting any photometric enhancement in the data will rely on either detecting an enhancement well above the natural variation expected due to rotation, or by averaging the photometric measurements over time, given that for a sufficient number of sparsely sampled ``snapshot'' observations (where full lightcurves are not obtained), rotational variations should ultimately average to zero.  To account for nights when at least partial lightcurves were obtained (i.e., where some photometric variation is clearly present), we compute the uncertainty, $\sigma_{m}$, for the average magnitude of each night's observations using
\begin{equation}
\sigma_{m} = {\Delta m_{\rm exp} - \Delta m_{\rm obs}\over 2}
\end{equation}
where $\Delta m_{\rm exp}$ is the expected or assumed total photometric range, assumed here to be $\Delta m_{\rm exp}=0.6$~mag, and $\Delta m_{\rm obs}$ is the observed photometric range.  We then use these magnitude uncertainties to compute weighted average magnitudes over the time periods in which we are interested (where in all cases, this assumed rotational uncertainty dominates the photometric uncertainties for all of our data).  This technique is similar to those used to compute phase functions for objects where rotational lightcurves are not known \citep[e.g.,][]{mac12}.

While Figure~\ref{fig_absmag} clearly illustrates the $\sim$0.3~mag photometric enhancement during 176P's 2005 active period that was noted by \citet{hsi11a}, we see no indication of a photometric enhancement in the average brightness of 176P as observed between June and September 2011 when it was traversing the same part of its orbit as in 2005 (marked by the gray highlighted region in the figure), suggesting that it was inactive during this time.  We therefore conclude that no photometric indication of repeated activity in 176P in 2011 is detected by our analysis.

\subsection{Azimuthal Tail Search}\label{section_azimuthal}

Since in 2005, 176P's activity was clearly dominated by its faint tail with very little evidence of coma, we conduct a search for similar activity in our 2011 data where we assume directed dust emission in the form of a tail.  To do so, we use the composite images constructed for our one-dimensional surface brightness profile analyses (Section~\ref{section_sbpanalysis}), except that instead of rotating the images such that star trails are horizontal in the image frame, we maintain a standard orientation for all images with North up and East left.  We then measure the average brightness of the sky surrounding the object within small rectangles ($\sim$100 pixels each) at position angles around the object's nucleus ranging from 0$^{\circ}$ to 360$^{\circ}$ (East of North) in 15$^{\circ}$ intervals.  For each image, the sizes and positions of these rectangles are selected to exclude as much of the nucleus's PSF as possible while also avoiding any visible nearby field stars or galaxies.  We then subtract the average sky background (measured from nearby regions of blank sky) and normalize the results to the peak brightness of the object itself.

\begin{figure}
\plotone{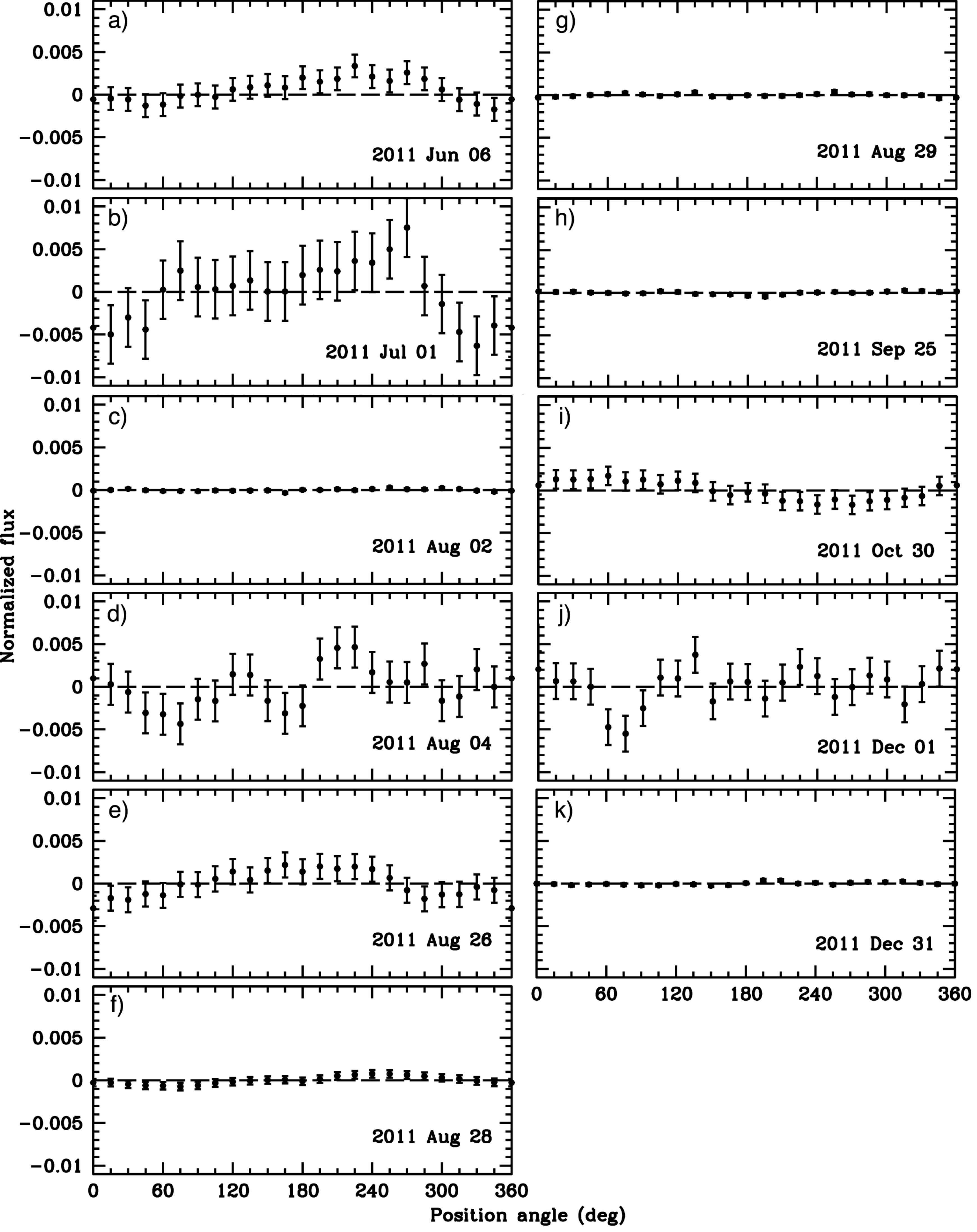}
\caption{\small 
Azimuthal surface brightness plots for composite images of 176P for (a) 2011 June 6, (b) 2011 July 1, (c) 2011 August 2, (d) 2011 August 4, (e) 2011 August 26, (f) 2011 August 28, (g) 2011 August 29, (h) 2011 September 25, (i) 2011 October 30, (j) 2011 December 1, and (k) 2011 December 31.
}
\label{fig_azplots}
\end{figure}

\begin{figure}
\plotone{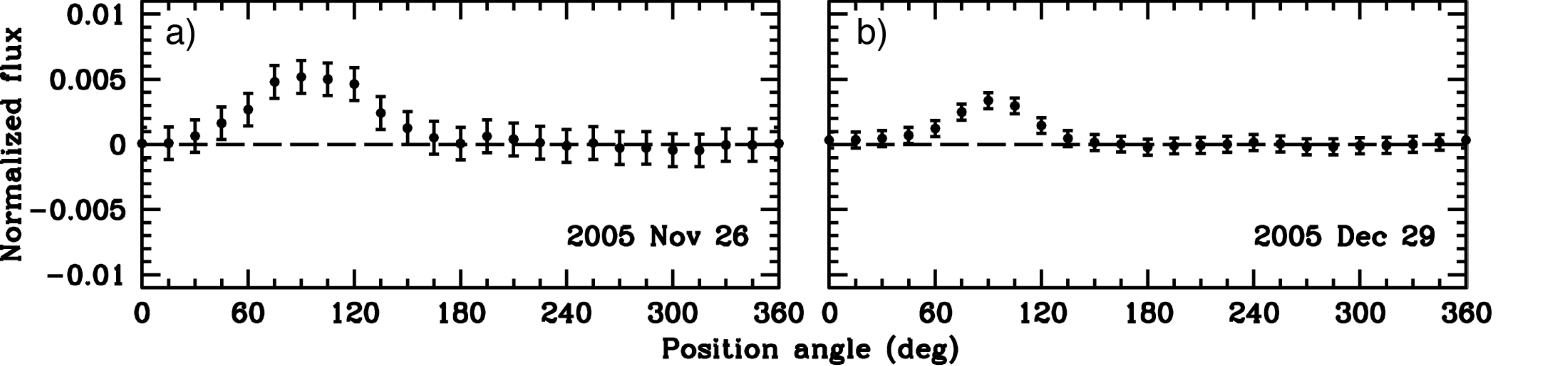}
\caption{\small 
Azimuthal surface brightness plots for composite images of 176P for (a) 2005 November 26, and (b) 2005 December 29.
}
\label{fig_azplots_active}
\end{figure}

Our data varies in quality over the course of our 2011 observations, as shown by the uncertainties plotted in Figure~\ref{fig_azplots}, which are the standard deviations computed for several samples of blank sky in each composite image.  Nonetheless, our resulting plots (Figure~\ref{fig_azplots}) show no consistent evidence of excess sky flux at a particular position angle around the nucleus (where the position angles of the projected anti-solar vector and orbit plane remain unchanged for the majority of the 2011 observing period) that would indicate the presence of a tail similar to that observed in 2005.  For comparison, the same analysis performed on the 2005 Gemini data for this object when it was active shows a clear and consistent brightness excess at a position angle of $\sim$90$^{\circ}$ (Figure~\ref{fig_azplots_active}).

\subsection{Surface Brightness Profile Analysis}\label{section_sbpanalysis}

Another method of searching for low-level activity is by searching for deviations in an object's surface brightness profile as compared to surface brightness profiles of nearby stellar sources.  Due to the apparent non-sidereal motion of 176P on the sky and our generally long ($>300$~s) exposure times, however, stellar sources are significantly trailed in the majority of our data.  PS1 data were obtained using sidereal telescope tracking, and so for those data, stellar sources are untrailed while the object is trailed.  Due to the short exposure times of the PS1 observations, the trailing of the object is smaller in those data compared to our other data, but given the low level of activity we wish to be sensitive to, we also consider that trailing to be non-negligible.  As such, to avoid these trailing effects, we conduct our search for activity in 176P by analyzing one-dimensional surface brightness profiles measured perpendicularly to the direction of the object's apparent non-sidereal motion (i.e., the direction of trailing) \citep[cf.][]{luu92,hsi05}.

\begin{figure}
\plotone{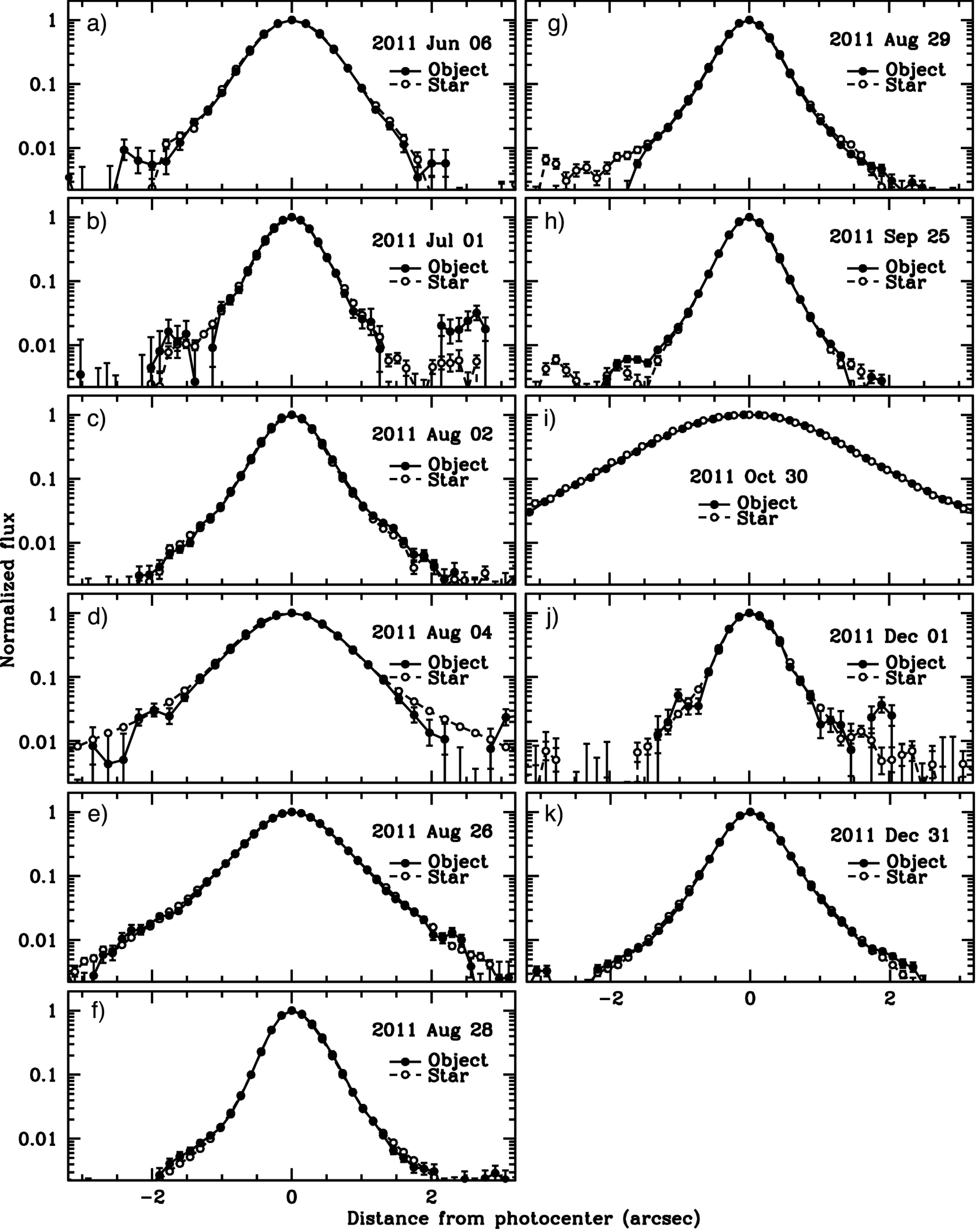}
\caption{\small 
One-dimensional surface brightness profiles plots for composite images of 176P overplotted on surface brightness profile plots of reference field stars for comparison for (a) 2011 June 6, (b) 2011 July 1, (c) 2011 August 2, (d) 2011 August 4, (e) 2011 August 26, (f) 2011 August 28, (g) 2011 August 29, (h) 2011 September 25, (i) 2011 October 30, (j) 2011 December 1, and (k) 2011 December 31.  Surface brightness is normalized to unity at each profile's peak and is plotted on a logarithmic scale versus angular distance in the plane of the sky.
}
\label{fig_sbp_plots}
\end{figure}

In addition to the composite images constructed by aligning images on 176P described in Section~\ref{section_observations}, we also construct composite images aligned on field stars.  To do so, instead of computing the appropriate offsets from a single reference point (i.e., the object), we compute average offsets from measurements of the photocenters of several nearby field stars.  All composite images (including those aligned on the object and on field stars) are then rotated by appropriate angles to align star (or object) trails horizontally in the image frame.  We then measure one-dimensional surface brightness profiles by averaging over horizontal rows over the entire widths of the object and reference stars and subtracting sky background sampled from nearby areas of blank sky.  Object and stellar profiles are then normalized to unity at their peaks and plotted together to search for dissimilarities, specifically excesses in 176P's profile, that could indicate the presence of near-nucleus coma (Figure~\ref{fig_sbp_plots}).

\begin{figure}
\plotone{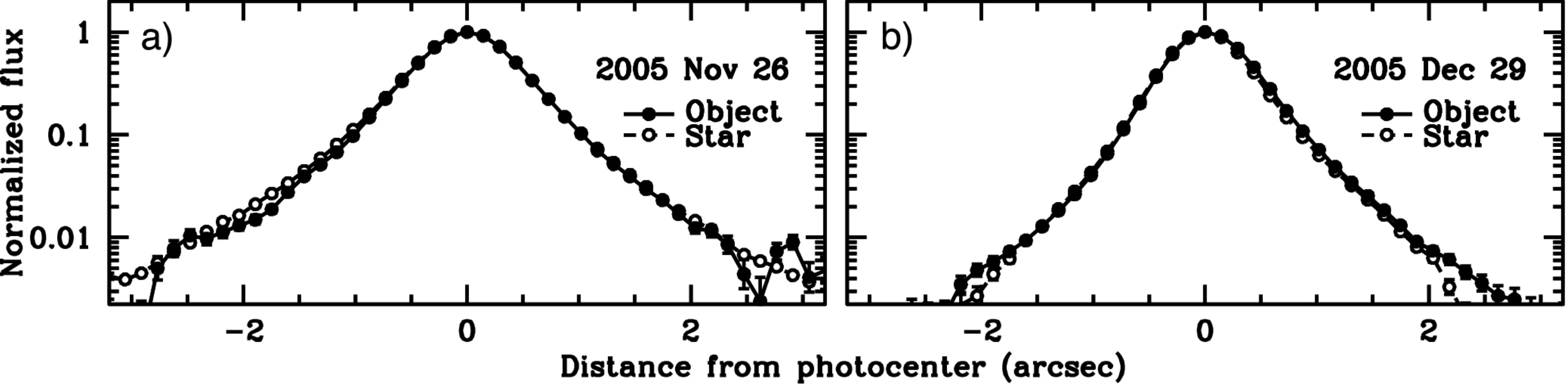}
\caption{\small 
One-dimensional surface brightness profiles plots for composite images of 176P overplotted on surface brightness profile plots of reference field stars for comparison for (a) 2005 November 26, and (b) 2005 December 29.  Surface brightness is normalized to unity at each profile's peak and is plotted on a logarithmic scale versus angular distance in the plane of the sky.
}
\label{fig_sbp_plots_active}
\end{figure}

We perform this procedure for all data obtained between June and December 2011 during which the object was expected to be potentially active, and find no consistent evidence of excess flux that would indicate the presence of a coma.  We note however that when it was initially observed to be active in 2005, 176P primarily exhibited a faint tail with little evidence of any coma.  As such, for comparison, we perform this same surface brightness analysis on Gemini observations of 176P from 2005 when the tail was clearly visible \citep{hsi11a}, and plot the results in Figure~\ref{fig_sbp_plots_active}.  No evidence of a coma is seen in these plots either, indicating that even if 176P had exhibited the same type of activity in 2011 as it did in 2005, this one-dimensional surface brightness analysis would not be the ideal means for detecting it, given that it is primarily sensitive to radially-symmetric coma and not directed emission.

\section{DISCUSSION}\label{section_discussion}

\citet{jew12} described several mechanisms by which a small solar system body could eject mass and potentially be observed to exhibit comet-like activity.  Among the mechanisms considered were rotational instability, electrostatic levitation, thermal fracturing, shock dehydration, and radiation pressure sweeping, but as many of these mechanisms require certain specific and often atypical physical conditions to be plausible, the primary dust ejection mechanisms considered when comet-like activity is observed are sublimation and impact ejection.

\citet{hsi12a} described several possible criteria for discriminating between these two primary mass loss mechanisms.  Among the conditions indicating that activity could be sublimation-driven are steadily increasing activity levels during long-lived active episodes \citep[e.g., as in the cases of P/La Sagra and P/2012 T1;][]{hsi12c,hsi13a} , ``comet-like'' dust cloud morphologies \citep[e.g., as in the cases of 259P and P/2006 VW$_{139}$;][]{jew09,hsi12b}, and recurrent activity separated by intervening periods of inactivity \citep[e.g., as in the cases of 133P and 238P;][]{hsi13,hsi04,hsi10,hsi11b}.  In contrast, among the criteria indicating that activity could be impact-generated (and that the body itself may not necessarily contain any ice) are rapidly decreasing activity levels and short-lived active episodes, unusual dust cloud morphologies, and a lack of repeated activity.  It was noted, however, that many of these criteria are not definitive indicators on their own and that supplemental evidence and analysis (e.g., numerical dust modeling) is generally needed to confidently determine the nature of a given comet-like object \citep[e.g.,][]{hsi04,hsi09a,hsi11a,hsi12b,hsi12c,jew10,jew11,sno10,ish11,mor11a,mor11b,mor12,mor13,ste12}.  Even then, due to the typically highly underconstrained nature of these cases and the large number of free parameters in all of the dust models in use, even dust modeling can sometimes lead to incorrect conclusions \citep[e.g.,][]{mor10}.

In the case of 176P, \citet{hsi11a} observationally confirmed the presence of persistent activity over at least 33 days (2005 November 26 through 2005 December 29), and perhaps as long as 66 days depending on whether an earlier photometric detection of activity on 2005 October 24 is considered reliable.  However, no significant increase or decrease of activity strength was clearly detected during this period.  In terms of morphology, 176P's activity primarily consisted of a faint fan-shaped dust tail with no significant coma.  Notably, the dust tail was not aligned either with the projection of the anti-solar vector on the plane of the sky, nor with the object's orbit plane, however, leading the authors to suggest that the tail could be due to a near-polar jet of directed ejected material.  Finally, while dust modeling of directed jet-like emission persisting over the observed active period of the object was able to reproduce the observed morphological evolution of the tail, modeling of impulsive emission events (i.e., as would be expected from an impact) was not able to match the observations.

The key constraint in this case was the persistence of the activity.  \citet{hsi11a} observed activity on 2005 November 26 and 2005 December 29, but saw no evidence of activity on 2006 February 3.  In the event of an impulsive emission event, if large particles were assumed to dominate the observed dust cloud, modeling indicated that the tail should have still been visible during the February observations.  However, if smaller particles were assumed to dominate the observed dust cloud, the modeled tail would dissipate by the time of the February observations, but in that case, the tail would also have been absent at the time of the December observations.  As such, an impulsive dust emission event was ruled out on the basis that no single particle size distribution could reproduce a tail that remained visible in November and December, but disappeared by February.

Despite these modeling results, however, the lack of a detection of repeated activity in 2011 now casts doubt on the classification of 176P as a ``true'' MBC (for which cometary activity is driven by sublimation).  The significance of repeated activity is that assuming that an object possesses a supply of volatile material and that it is not completely exhausted from the object's previous active episode, that volatile material would be expected to continue to undergo future periods of sublimation, e.g., each time the object passes perihelion, such as in the case of 133P \citep{hsi04,hsi10,low05,hsi13}.  In the case of an impact-generated active episode, however, repeated activity over a short period of time would not be expected since that would require the same object to experience multiple impacts in that span of time when similar impacts are not observed at anywhere near the same frequency for other asteroids.

Could 176P's activity simply have escaped detection in 2011?  While the heliocentric distance of the object during its 2011 perihelion passage was similar to its heliocentric distance in 2005, the geocentric distance ranged between $\Delta=1.6$~AU and $\Delta=2.2$~AU ($\nu=1.4^{\circ}$ to $\nu=18.6^{\circ}$) during its 2005 active period.  For comparison, in 2011, over the same true anomaly range, 176P's geocentric distance ranged from $\Delta=2.9$~AU to $\Delta=2.2$~AU.  As such, even though 176P's discovery images were taken at a much closer geocentric distance than any observations taken in 2011, we at least have directly comparable sets of observations from 2005 December 29 (Gemini, 3060~s effective exposure time, $\nu=18.6^{\circ}$, $R=2.60$~AU, $\Delta=2.2$~AU, $\alpha=21.7^{\circ}$, FWHM seeing $\sim$ $0\farcs7$), when activity was clearly visible, and 2011 September 25 (Gemini, 1440~s effective exposure time, $\nu=22.4^{\circ}$, $R=2.61$~AU, $\Delta=1.99$~AU, $\alpha=19.9^{\circ}$, FWHM seeing $\sim$ $0\farcs5$), when no activity was detected.  Revisiting the 2005 data, we find a tail surface brightness of $\Sigma=25.2$~mag~arcsec$^{-2}$ in the 2005 December 29 Gemini data.  Such a tail, if present, should have been clearly detectable in the 2011 September 25 Gemini data for which we compute a 3-$\sigma$ surface brightness detection limit of $\Sigma_{\rm lim}=27.4$~mag~arcsec$^{-2}$.  Given that no such tail is evident at all in the data (Figures~\ref{fig_images_2}h, \ref{fig_azplots}h), we conclude that no activity is present.  This result confirms the conclusions of \citet{dev12} that any activity in 2011 was substantially weaker than in 2005, as well as our photometric analysis showing that the object's brightness in 2011 is consistent with brightness predictions for an inactive nucleus with no evidence of the brightness enhancement observed during the object's active period in 2005.

The non-detection of activity during 176P's 2011 perihelion passage appears to suggest that its 2005 activity could have been due to an impact and that the object should actually be considered a disrupted asteroid and not a MBC.  As discussed above, in 2005, 176P's tail did not correspond to either the direction of the projected antisolar vector on the sky or with the object's orbit plane \citep{hsi11a}, similar to what was observed for disrupted asteroid (596) Scheila in 2010, suggesting that an unusual dust ejection mechanism, such as a collision, could be responsible.  The relatively constant activity level observed for 176P over its 2005 active period stands in sharp contrast, however, to Scheila's rapid fading \citep[30\% in just 8 days;][]{jew11}.

Rather than exhibiting impact-generated dust emission, then, it could simply be that 176P did in fact exhibit sublimation-driven activity in 2005, but in doing so, exhausted most of its limited supply of volatile material.  Mantling could have also occurred following the initial active episode \citep[cf.][]{jew96}, quenching future outbursts.  Both of these possibilities would be consistent with the conclusions of \citet{hsi11a} who suggested that 176P was primarily ejecting dust from an isolated active site near one of its rotational poles (leading to jet-like directed emission), since the smaller the total active area on 176P's surface responsible for driving its activity, the more plausible it would be for the volatile supply near that particular site to be either effectively completely exhausted or effectively completely quenched by mantling by a single active episode.  The greater longevity of activity on other MBCs like 133P and 238P, both in terms of the duration of individual active episodes and the number of times those active episodes repeat, could be due to those objects having a larger total volatile content, or alternatively simply having larger or more volatile-rich active sites.

Unfortunately, it is unclear how we might be able to observationally confirm this hypothesis.  If the active site responsible for 176P's 2005 activity is now completely devolatilized or quenched by mantling, no further activity would be expected.  This expectation of course would be the same if 176P's original activity were due to an impact, and so is not a particularly discriminating prediction.  The only meaningful path forward may simply be to conduct further characterization studies of other MBCs, focusing in particular on monitoring changes in activity strength from one active episode to another to better understand volatile depletion and mantling processes on MBCs and gain insights that could be useful for explaining 176P's behavior.  Continued monitoring of 176P to search for resumed activity in the future could also be useful in the event that a new active site is exposed either by another collision or thermal stresses.

% activity may be even more difficult to observe during 176P's next perihelion passage in March 2017 as the object will not become visible from the Earth after passing perihelion until July, by which time it will have already reached a true anomaly of $\nu\sim30^{\circ}$ and will have a geocentric distance of $\Delta\sim3.4$~AU.  Observations will be possible however in January 2017 when the object will be at a true anomaly of $\nu\sim350^{\circ}$ and a geocentric distance of $\Delta=3.2$~AU.

\section{SUMMARY}\label{section_summary}

We present the results of a search for the reactivation of 176P during its 2011 perihelion passage using deep optical observations obtained before, during, and after that perihelion passage, finding the following key results:

\begin{enumerate}
\item{Photometric measurements of 176P obtained between June 2011 and December 2011 show no evidence of a photometric enhancement similar to that observed for 176P between November 2005 and December 2005 when it was previously observed to be visibly active.  The average magnitude of the object measured during this period is entirely consistent with brightness predictions for an inactive nucleus.}
\item{An azimuthal search for directed dust emission (e.g., a tail) likewise revealed no consistent evidence for activity in 2011 in the form of directed emission, though a similar analysis of data obtained in 2005 show clear evidence for such directed emission.}
\item{We also conduct a one-dimensional surface brightness profile analysis for data obtained in 2011, finding no evidence of activity in the form of excess flux in the surface brightness profile of the object as compared to nearby reference field stars.  We note however that a similar analysis of data obtained in 2005 also show no evidence of excess flux, even though activity is clearly present in the form of a tail, and attribute this finding to the fact that this technique is best suited for the detection of a spherically symmetric coma and not directed emission, which is what was previously observed for 176P.}
\item{Finally, we find that a comparison of observational circumstances of data sets obtained in 2005 and in 2011 imply that activity similar to that exhibited by 176P in 2005 should have also have been detectable in 2011 with the data presented here.  Given the lack of detection of any activity using any of our other quantitative means, we conclude that our non-detection of activity during 176P's 2011 perihelion passage is real.}
\item{The lack of repeated activity in 176P in 2011 casts doubt on the sublimation-driven nature of the activity observed in 2005, but does not completely rule it out, particularly since other lines of evidence continue to support the possibility that the 2005 activity was cometary in nature.  We speculate that the active site on 176P responsible for its activity in 2005 could have been depleted of volatiles or possibly quenched by mantling, both of which would cause the cessation of any future active episodes.  Observations of other MBCs over the course of successive active episodes to characterize the decay of activity strength should contribute to a better understanding of volatile depletion and mantling processes on MBCs that could be relevant to understanding 176P's behavior.}
\end{enumerate}

\begin{acknowledgements}
H.H.H.\ acknowledges support for this work by NASA through Hubble Fellowship grant HF-51274.01 awarded by the Space Telescope Science Institute, which is operated by the Association of Universities for Research in Astronomy (AURA) for NASA, under contract NAS 5-26555, as well as by the United KingdomÕs Science and Technology Facilities Council (STFC) through STFC fellowship grant ST/F011016/1.
H.M.K., K.J.M., and B.Y. acknowledge support through the NASA Astrobiology Institute under Cooperative Agreement NNA09DA77A, while
J.K. acknowledges support through NSF grant AST 1010059 and 
C.S. acknowledges support from the European Union Seventh Framework Programme (FP7/2007-2013) under grant agreement no.\ 268421.
%B.N.\ is supported by the Ministry of Education and Science of Serbia (Project 176011).
%A.F.\ is supported by the Science \& Technology Facilities Council (Grant ST/F002270/1).
%M.S.K.\ is supported by NASA Planetary Astronomy Grant NNX09AF10G.
%We thank John Dvorak, Callie McNew, and Erik Moore at the UH 2.2~m telescope, XXX other TO acknowledgements XXX, Luca Rizzi, Greg Wirth, Heather Hershley, and Julie Renaud-Kim at Keck, Ang\'elica Le\'on at the NTT for their assistance in obtaining observations.
The Pan-STARRS1 Surveys (PS1) have been made possible through contributions of the Institute for Astronomy, the University of Hawaii, the Pan-STARRS Project Office, the Max-Planck Society and its participating institutes, the Max Planck Institute for Astronomy, Heidelberg and the Max Planck Institute for Extraterrestrial Physics, Garching, The Johns Hopkins University, Durham University, the University of Edinburgh, Queen's University Belfast, the Harvard-Smithsonian Center for Astrophysics, the Las Cumbres Observatory Global Telescope Network Incorporated, the National Central University of Taiwan, the Space Telescope Science Institute, the National Aeronautics and Space Administration under Grant No.\ NNX08AR22G issued through the Planetary Science Division of the NASA Science Mission Directorate, the National Science Foundation under Grant No.\ AST-1238877, the University of Maryland, and Eotvos Lorand University (ELTE).
We thank the PS1 Builders and PS1 operations staff for construction and operation of the PS1 system and access to the data products provided.
Gemini Observatory is operated by the Association of Universities for Research in Astronomy, Inc., under a cooperative agreement with the NSF on behalf of the Gemini partnership: the National Science Foundation (United States), the National Research Council (Canada), CONICYT (Chile), the Australian Research Council (Australia), Minist\'{e}rio da Ci\^{e}ncia, Tecnologia e Inova\c{c}\~{a}o (Brazil) and Ministerio de Ciencia, Tecnolog\'{i}a e Innovaci\'{o}n Productiva (Argentina).
%The WHT is operated by the Isaac Newton Group in the Observatorio del Roque de los Muchachos of the Instituto de Astrof\'isica de Canarias.
%The Faulkes Telescopes are operated by Las Cumbres Observatory Global Telescope Network.
%SDSS-III ({\tt http://www.sdss3.org/}) is funded by the Alfred P. Sloan Foundation, the Participating Institutions, NSF, and the U.S.\ Department of Energy Office of Science, and managed by the Astrophysical Research Consortium for the SDSS-III Collaboration.
%The PS1 Surveys have been made possible through contributions of the Institute for Astronomy, the University of Hawaii, the Pan-STARRS Project Office, the Max-Planck Society and its participating institutes, the Max Planck Institute for Astronomy, Heidelberg and the Max Planck Institute for Extraterrestrial Physics, Garching, The Johns Hopkins University, Durham University, the University of Edinburgh, Queen's University Belfast, the Harvard-Smithsonian Center for Astrophysics, and the Las Cumbres Observatory Global Telescope Network, Incorporated, the National Central University of Taiwan, and the National Aeronautics and Space Administration under Grant No. NNX08AR22G issued through the Planetary Science Division of the NASA Science Mission Directorate.
\end{acknowledgements}

\begin{deluxetable}{lcrrcrrrrrrrcc}
%\footnotesize
\scriptsize
\tablewidth{0pt}
\tablecaption{New Observations of 176P/LINEAR\label{table_obslog_new}}
\tablecolumns{14}
\tablehead{
\colhead{UT Date}
 & \colhead{Tel.\tablenotemark{a}}
 & \colhead{N\tablenotemark{b}}
 & \colhead{t\tablenotemark{c}}
 & \colhead{Filter}
 & \colhead{$\nu$\tablenotemark{d}}
 & \colhead{$R$\tablenotemark{e}}
 & \colhead{$\Delta$\tablenotemark{f}}
 & \colhead{$\alpha$\tablenotemark{g}}
 & \colhead{$\alpha_{pl}$\tablenotemark{h}}
 & \colhead{PA$_{-\odot}$\tablenotemark{i}}
 & \colhead{PA$_{-v}$\tablenotemark{j}}
 & \colhead{$m_R(R,\Delta,\alpha)$\tablenotemark{k}}
 & \colhead{$m_R(1,1,0)$\tablenotemark{l}}
}
\startdata
2005 Oct 18    & \multicolumn{4}{l}{\it Perihelion..........................} & 0.0 & 2.581 & 1.586 & 1.4 & $-$0.1 & 71.5 & 247.8 & --- & --- \\
%2010 Aug 1-2   & LT     &    &  & $BR$           & 2.962 & 1.955 &  3.0 &    0.1 &  75.9 & 257.4 & 281.2 \\
%2010 Aug 2-3   & LT     &    &  & $BR$           & 2.960 & 1.956 &  3.4 &    0.1 &  76.2 & 257.5 & 281.4 \\
%2010 Aug 6-7   & LT     &    &  & $BR$           & 2.953 & 1.961 &  3.0 &    0.1 &  76.9 & 257.7 & 282.2 \\
%2010 Aug 8-9   & LT     &    &  & $BR$           & 2.949 & 1.964 &  5.7 &    0.1 &  77.2 & 257.9 & 282.6 \\
%2010 Aug 9-10  & LT     &    &  & $BR$           & 2.947 & 1.967 &  6.1 &    0.1 &  77.3 & 257.9 & 282.8 \\
%2010 Aug 11-12 & LT     &    &  & $BR$           & 2.944 & 1.972 &  6.8 &    0.1 &  77.6 & 258.1 & 283.2 \\
%2010 Aug 12-13 & LT     &    &  & $BR$           & 2.942 & 1.975 &  7.2 &    0.1 &  77.7 & 258.1 & 283.4 \\
%2010 Aug 14-15 & LT     &    &  & $BR$           & 2.938 & 1.982 &  7.9 &    0.1 &  77.9 & 258.2 & 283.8 \\
%2010 Aug 15-16 & LT     &    &  & $BR$           & 2.936 & 1.986 &  8.3 &    0.1 &  77.9 & 258.3 & 284.0 \\
%2010 Aug 17-18 & LT     &    &  & $BR$           & 2.933 & 1.994 &  9.0 &    0.0 &  78.1 & 258.4 & 284.4 \\
%2010 Aug 18-19 & LT     &    &  & $BR$           & 2.931 & 2.000 &  9.4 &    0.0 &  78.2 & 258.5 & 284.6 \\
%2010 Aug 19-20 & LT     &    &  & $BR$           & 2.929 & 2.004 &  9.7 &    0.0 &  78.3 & 258.5 & 284.8 \\
%2010 Aug 23-24 & LT     &    &  & $BR$           & 2.922 & 2.025 & 11.0 &    0.0 &  78.5 & 258.7 & 285.7 \\
%2010 Aug 24-25 & LT     &    &  & $BR$           & 2.920 & 2.031 & 11.3 &    0.0 &  78.6 & 258.8 & 285.9 \\
2010 Aug 05    & NTT     & 54 & 16200 & $R$      & 281.9 & 2.956 & 1.958 &  4.3 &    0.1 &  76.7 & 257.6 & 19.45$\pm$0.03 & 15.26$\pm$0.30 \\
2010 Aug 06    & NTT     & 13 &  3900 & $R$      & 282.1 & 2.954 & 1.959 &  4.7 &    0.1 &  76.9 & 257.7 & 19.29$\pm$0.04 & 15.07$\pm$0.30 \\
2010 Aug 11    & UH2.2   & 14 &  4200 & $R$      & 283.0 & 2.945 & 1.970 &  6.6 &    0.1 &  77.5 & 258.0 & 19.21$\pm$0.02 & 14.89$\pm$0.20 \\
2010 Aug 13    & NTT     & 45 & 13500 & $R$      & 283.5 & 2.942 & 1.976 &  7.3 &    0.1 &  77.7 & 258.1 & 19.55$\pm$0.03 & 15.19$\pm$0.30 \\
2010 Aug 13    & UH2.2   & 26 &  7800 & $R$      & 283.5 & 2.941 & 1.977 &  7.4 &    0.1 &  77.7 & 258.1 & 19.45$\pm$0.02 & 15.09$\pm$0.15 \\
2010 Aug 14    & NTT     & 13 &  3900 & $R$      & 283.7 & 2.939 & 1.980 &  7.7 &    0.1 &  77.8 & 258.2 & 19.45$\pm$0.03 & 15.08$\pm$0.30 \\
2010 Aug 15    & NTT     & 28 &  8400 & $R$      & 283.9 & 2.938 & 1.984 &  8.1 &    0.1 &  77.9 & 258.3 & 19.44$\pm$0.03 & 15.05$\pm$0.30 \\
2010 Aug 16    & UH2.2   & 23 &  6900 & $R$      & 284.1 & 2.936 & 1.988 &  8.5 &    0.0 &  78.0 & 258.3 & 19.49$\pm$0.02 & 15.08$\pm$0.15 \\
2010 Aug 28    & du Pont & 11 &  1320 & $R$      & 286.5 & 2.914 & 2.051 & 12.3 &    0.0 &  78.8 & 258.9 & 19.79$\pm$0.03 & 15.17$\pm$0.10 \\
2010 Aug 29    & du Pont &  9 &  1080 & $R$      & 286.7 & 2.912 & 2.058 & 12.6 &    0.0 &  78.8 & 258.9 & 19.75$\pm$0.03 & 15.12$\pm$0.10 \\
2010 Aug 30    & du Pont &  6 &   720 & $R$      & 286.9 & 2.911 & 2.065 & 12.9 &    0.0 &  78.9 & 259.0 & 19.54$\pm$0.02 & 14.89$\pm$0.30 \\
2010 Aug 31    & du Pont &  5 &   600 & $R$      & 287.1 & 2.909 & 2.072 & 13.2 &    0.0 &  78.9 & 259.0 & 19.61$\pm$0.02 & 14.94$\pm$0.30 \\
2010 Sep 01    & UH2.2   & 16 &  4800 & $R$      & 287.4 & 2.906 & 2.081 & 13.6 &    0.0 &  78.9 & 259.0 & 19.96$\pm$0.02 & 15.27$\pm$0.15 \\
2010 Sep 04    & NTT     &  2 &   600 & $R$      & 287.9 & 2.901 & 2.103 & 14.3 &    0.0 &  79.0 & 259.1 & 19.77$\pm$0.04 & 15.04$\pm$0.30 \\
2010 Sep 05    & NTT     &  2 &   600 & $R$      & 288.1 & 2.900 & 2.111 & 14.6 &    0.0 &  79.1 & 259.1 & 20.20$\pm$0.03 & 15.45$\pm$0.30 \\
2010 Oct 05    & Keck    &  2 &   240 & $R$      & 294.5 & 2.846 & 2.415 & 19.8 &    0.0 &  78.8 & 258.7 & 20.48$\pm$0.04 & 15.30$\pm$0.30 \\
2010 Oct 23    & VLT     &  4 &   480 & $R$      & 298.3 & 2.816 & 2.624 & 20.7 & $-$0.1 &  77.8 & 257.6 & 20.81$\pm$0.16 & 15.45$\pm$0.30 \\
%2010 Nov 26    & UH2.2  &  4 &  1200 & $R$      & 306.0 & 2.761 & 3.019 & 19.0 & $-$0.1 &  74.7 & 254.4 & 20.84$\pm$0.12 & 15.27$\pm$0.30 \\ %nearby BS; SBPs no good
2011 Jun 06    & Subaru  &  7 &  1260 & $R$      & 353.6 & 2.579 & 3.213 & 15.8 &    0.1 & 249.8 & 249.5 & 20.27$\pm$0.02 & 14.81$\pm$0.30 \\
2011 Jul 01    & \multicolumn{4}{l}{\it Perihelion..........................} & 0.0 & 2.576 & 2.980 & 19.4 & 0.1 & 252.1 & 251.8 & --- & --- \\
2011 Jul 01    & VLT     &  4 &   480 & $R$      &   0.0 & 2.576 & 2.976 & 19.4 &    0.1 & 252.1 & 251.9 & 20.64$\pm$0.06 & 15.24$\pm$0.30 \\
2011 Aug 02    & Gemini  &  6 &  1800 & $r'$     &   8.5 & 2.581 & 2.617 & 22.5 &    0.1 & 255.4 & 255.3 & 20.01$\pm$0.02 & 14.79$\pm$0.30 \\
2011 Aug 04    & UH2.2   & 13 &  3900 & $R$      &   9.0 & 2.582 & 2.594 & 22.6 &    0.1 & 255.6 & 255.5 & 19.99$\pm$0.02 & 14.78$\pm$0.30 \\
2011 Aug 26    & Keck    & 11 &  1320 & $R$      &  14.7 & 2.590 & 2.330 & 22.9 &    0.0 & 257.6 & 257.6 & 19.92$\pm$0.02 & 14.93$\pm$0.30 \\
2011 Aug 28    & Gemini  &  3 &   900 & $r'$     &  15.3 & 2.591 & 2.306 & 22.9 &    0.0 & 257.8 & 257.7 & 20.07$\pm$0.02 & 15.10$\pm$0.30 \\
2011 Aug 29    & Gemini  &  6 &  1800 & $r'$     &  15.5 & 2.592 & 2.294 & 22.8 &    0.0 & 257.9 & 257.8 & 20.15$\pm$0.02 & 15.19$\pm$0.30 \\
2011 Sep 25    & Gemini  &  8 &  1440 & $r'$     &  22.4 & 2.608 & 1.987 & 19.9 &    0.0 & 259.2 & 259.3 & 19.94$\pm$0.02 & 15.37$\pm$0.30 \\
%2011 Oct 28    & UH2.2  &    &       &          &  30.8 & 2.637 & 1.720 & 10.4 & $-$0.1 & 258.3 & 258.8 &  \\ %no SDSS or Landolt calibration available
2011 Oct 30    & UH2.2   & 12 &  3600 & $R$      &  31.3 & 2.639 & 1.711 &  9.6 & $-$0.1 & 258.1 & 258.7 & 18.82$\pm$0.02 & 14.92$\pm$0.30 \\
%2011 Nov 24    & PS1    &  2 &       & $r_{P1}$ &  37.5 & 2.665 & 1.679 &  1.3 & $-$0.1 &  82.7 & 257.0 &  \\
2011 Dec 01    & PS1     &  2 &    80 & $r_{P1}$ &  39.2 & 2.674 & 1.701 &  4.4 & $-$0.1 &  78.2 & 256.5 & 18.86$\pm$0.05 & 15.18$\pm$0.30 \\
2011 Dec 22    & NTT     &  5 &  1500 & $R$      &  44.1 & 2.700 & 1.842 & 12.4 & $-$0.1 &  76.0 & 255.4 & 19.29$\pm$0.04 & 15.07$\pm$0.30 \\
2011 Dec 31    & Gemini  &  9 &  1620 & $r'$     &  46.4 & 2.712 & 1.936 & 15.1 & $-$0.1 &  75.7 & 255.2 & 19.59$\pm$0.02 & 15.15$\pm$0.30 \\
2012 Nov 13    & UH2.2   &  8 &  2400 & $R$      & 108.8 & 3.278 & 3.239 & 17.5 & $-$0.1 & 290.2 & 290.4 & 21.25$\pm$0.10 & 15.20$\pm$0.30 \\
2012 Dec 18    & UH2.2   &  4 &  1200 & $R$      & 114.4 & 3.340 & 2.797 & 15.4 & $-$0.1 & 290.8 & 291.1 & 20.60$\pm$0.06 & 14.90$\pm$0.30 \\
2013 May 12    & UH2.2   &  6 &  1800 & $R$      & 135.4 & 3.564 & 3.436 & 16.5 &    0.1 & 108.5 & 288.7 & 21.23$\pm$0.08 & 14.91$\pm$0.30 \\
2013 May 13    & UH2.2   &  3 &   900 & $R$      & 135.5 & 3.565 & 3.452 & 16.5 &    0.1 & 108.5 & 288.8 & 21.19$\pm$0.08 & 14.86$\pm$0.30 \\
2017 Mar 12    & \multicolumn{4}{l}{\it Perihelion..........................} & 0.0 & 2.580 & 3.476 & 8.2 & 0.0 & 66.8 & 246.9 & --- & ---
\enddata
\tablenotetext{a}{Telescope used.}
\tablenotetext{b}{Number of exposures.}
\tablenotetext{c}{Total integration time, in s.}
\tablenotetext{d}{True anomaly, in degrees.}
\tablenotetext{e}{Heliocentric distance of object, in AU.}
\tablenotetext{f}{Geocentric distance of object, in AU.}
\tablenotetext{g}{Solar phase angle (Sun-object-Earth), in degrees.}
\tablenotetext{h}{Orbit plane angle, in degrees.}
\tablenotetext{i}{Position angle of the antisolar vector, in degrees East of North.}
\tablenotetext{j}{Position angle of the negative velocity vector, in degrees East of North.}
\tablenotetext{k}{Apparent $R$-band magnitude.}
\tablenotetext{l}{Absolute $R$-band magnitude at $R=\Delta=1$~AU and $\alpha=0^{\circ}$, computed using phase function parameters from \citet{hsi09b}.}
\end{deluxetable}

\begin{deluxetable}{lcrrcrrcc}
%\footnotesize
\scriptsize
\tablewidth{0pt}
\tablecaption{Previous Observations of 176P/LINEAR\label{table_obslog_old}}
\tablecolumns{9}
\tablehead{
\colhead{UT Date}
 & \colhead{Tel.\tablenotemark{a}}
 & \colhead{t\tablenotemark{b}}
 & \colhead{$\nu$\tablenotemark{c}}
 & \colhead{$R$\tablenotemark{d}}
 & \colhead{$\Delta$\tablenotemark{e}}
 & \colhead{$\alpha$\tablenotemark{f}}
 & \colhead{$m_R(R,\Delta,\alpha)$\tablenotemark{g}}
 & \colhead{$m_R(1,1,0)$\tablenotemark{h}}
}
\startdata
2005 Oct 18 & \multicolumn{2}{l}{\it Perihelion ........} & 0.0 & 2.581 & 1.586 & 1.4 & --- & --- \\
2005 Oct 24 & Lulin  &  3000 &   1.4 & 2.581 & 1.598 & 4.2 &  18.20$\pm$0.01 & 14.74$\pm$0.30 \\
2005 Nov 26 & Gemini &   240 &  10.1 & 2.588 & 1.817 & 16.3 & 19.11$\pm$0.04 & 14.91$\pm$0.30 \\
2005 Dec 22 & UH2.2  &  2100 &  16.8 & 2.599 & 2.121 & 21.1 & 19.65$\pm$0.01 & 14.91$\pm$0.30 \\
2005 Dec 24 & UH2.2  &  7800 &  17.3 & 2.600 & 2.147 & 21.3 & 19.46$\pm$0.01 & 14.65$\pm$0.30 \\
2005 Dec 25 & UH2.2  &  9900 &  17.5 & 2.601 & 2.161 & 21.4 & 19.62$\pm$0.01 & 14.79$\pm$0.30 \\
2005 Dec 26 & UH2.2  &  9300 &  17.8 & 2.601 & 2.174 & 21.5 & 19.62$\pm$0.01 & 14.80$\pm$0.30 \\
2005 Dec 27 & UH2.2  &  8700 &  18.0 & 2.602 & 2.187 & 21.5 & 19.59$\pm$0.01 & 14.76$\pm$0.30 \\
2005 Dec 29 & Gemini &  3060 &  18.6 & 2.603 & 2.214 & 21.7 & 19.62$\pm$0.01 & 14.77$\pm$0.30 \\
2006 Feb 03 & UH2.2  &  3300 &  27.7 & 2.630 & 2.707 & 21.2 & 20.25$\pm$0.01 & 14.95$\pm$0.30 \\ %nonphot
2006 Aug 31 & UH2.2  &   600 &  75.2 & 2.933 & 3.546 & 14.3 & 21.23$\pm$0.08 & 15.23$\pm$0.30 \\ %nonphot
2006 Sep 02 & UH2.2  &   600 &  75.6 & 2.937 & 3.530 & 14.6 & 21.09$\pm$0.05 & 15.16$\pm$0.30 \\ %nonphot
2006 Dec 11 & UH2.2  &  9000 &  94.5 & 3.124 & 2.424 & 14.5 & 20.01$\pm$0.01 & 14.81$\pm$0.30 \\
2006 Dec 16 & UH2.2  &  3300 &  95.3 & 3.133 & 2.378 & 13.3 & 20.13$\pm$0.01 & 14.97$\pm$0.30 \\ %nonphot
2006 Dec 18 & UH2.2  &   900 &  95.7 & 3.137 & 2.361 & 12.8 & 20.09$\pm$0.09 & 14.99$\pm$0.30 \\ %nonphot
2007 Jan 27 & Keck   &   240 & 102.6 & 3.211 & 2.227 &  0.8 & 19.50$\pm$0.01 & 15.09$\pm$0.30 \\
2007 Feb 15 & UH2.2  &  9000 & 105.7 & 3.246 & 2.326 &  7.5 & 19.87$\pm$0.01 & 15.01$\pm$0.30 \\
2007 Feb 16 & UH2.2  & 14700 & 105.9 & 3.248 & 2.334 &  7.8 & 19.93$\pm$0.01 & 15.05$\pm$0.30 \\
2007 Mar 21 & UH2.2  & 10500 & 111.2 & 3.307 & 2.718 & 15.4 & 20.71$\pm$0.01 & 15.08$\pm$0.30 \\
2007 Mar 22 & UH2.2  & 15300 & 111.3 & 3.309 & 2.732 & 15.5 & 20.82$\pm$0.01 & 15.20$\pm$0.30 \\
2007 May 19 & UH2.2  &  2100 & 120.2 & 3.407 & 3.637 & 16.1 & 21.57$\pm$0.05 & 15.23$\pm$0.30 \\
2008 Jun 29 & NTT    &   360 & 173.2 & 3.803 & 3.795 & 15.4 & 21.68$\pm$0.07 & 14.95$\pm$0.30 \\ %nonphot
2008 Jun 30 & NTT    &   540 & 173.3 & 3.804 & 3.810 & 15.3 & 21.70$\pm$0.05 & 15.05$\pm$0.30 \\
2008 Jul 01 & NTT    &   540 & 173.4 & 3.804 & 3.824 & 15.3 & 21.63$\pm$0.05 & 14.97$\pm$0.30 \\
%2008 Aug 25 & \multicolumn{2}{l}{{\it Aphelion} ........... } & 180.0 & 3.810 & 4.523 &  9.9 & --- & --- \\
2009 Jan 23 & WHT    &   240 & 198.1 & 3.765 & 4.012 & 14.1 & 21.47$\pm$0.10 & 14.77$\pm$0.30 \\
2009 May 03 & INT    &   600 & 210.6 & 3.687 & 2.702 &  3.9 & 20.30$\pm$0.04 & 14.95$\pm$0.30 \\
2011 Jul 01    & \multicolumn{2}{l}{\it Perihelion ........} & 0.0 & 2.576 & 2.980 & 19.4 & --- & ---
\enddata
\tablenotetext{a}{Telescope used.}
\tablenotetext{b}{Total integration time, in s.}
\tablenotetext{c}{True anomaly, in degrees.}
\tablenotetext{d}{Heliocentric distance of object, in AU.}
\tablenotetext{e}{Geocentric distance of object, in AU.}
\tablenotetext{f}{Solar phase angle (Sun-object-Earth), in degrees.}
\tablenotetext{g}{Apparent $R$-band magnitude, as reported in \citet{hsi11a}.}
\tablenotetext{h}{Absolute $R$-band magnitude at $R=\Delta=1$~AU and $\alpha=0^{\circ}$, computed using phase function parameters from \citet{hsi09b}.}
\end{deluxetable}

%\begin{figure}
%\plotone{fig_sbp_models.pdf}
%\caption{\small 
%One-dimensional surface brightness profiles plots for composite images of 176P overplotted on surface brightness profile plots of comet models for comparison for (a) 2011 June 06, and (b) 2011 August 02.  Surface brightness is normalized to unity at each profile's peak and is plotted on a logarithmic scale versus angular distance in the plane of the sky.  Surface brightness profiles of 176P are indicated by blue lines, while surface brightness profiles of comet models with varying levels of coma ($\eta$) are indicated by dark gray, medium gray, and light gray lines.
%}
%\label{fig_sbp_models}
%\end{figure}

\end{document}